\PassOptionsToPackage{unicode}{hyperref}
\PassOptionsToPackage{hyphens}{url}
\PassOptionsToPackage{dvipsnames,svgnames,x11names}{xcolor}
\documentclass[12pt]{article}
\usepackage{amsmath,amssymb}
\usepackage{iftex}
\ifPDFTeX
  \usepackage[T1]{fontenc}
  \usepackage[utf8]{inputenc}
  \usepackage{textcomp} 
\else 
  \usepackage{unicode-math}
  \defaultfontfeatures{Scale=MatchLowercase}
  \defaultfontfeatures[\rmfamily]{Ligatures=TeX,Scale=1}
\fi
\usepackage{lmodern}
\ifPDFTeX\else  
\fi
\IfFileExists{upquote.sty}{\usepackage{upquote}}{}
\IfFileExists{microtype.sty}{
  \usepackage[]{microtype}
  \UseMicrotypeSet[protrusion]{basicmath} 
}{}
\makeatletter
\@ifundefined{KOMAClassName}{
  \IfFileExists{parskip.sty}{%
    \usepackage{parskip}
  }{
    \setlength{\parindent}{0pt}
    \setlength{\parskip}{6pt plus 2pt minus 1pt}}
}{
  \KOMAoptions{parskip=half}}
\makeatother
\usepackage{xcolor}
\makeatletter
\ifx\paragraph\undefined\else
  \let\oldparagraph\paragraph
  \renewcommand{\paragraph}{
    \@ifstar
      \xxxParagraphStar
      \xxxParagraphNoStar
  }
  \newcommand{\xxxParagraphStar}[1]{\oldparagraph*{#1}\mbox{}}
  \newcommand{\xxxParagraphNoStar}[1]{\oldparagraph{#1}\mbox{}}
\fi
\ifx\subparagraph\undefined\else
  \let\oldsubparagraph\subparagraph
  \renewcommand{\subparagraph}{
    \@ifstar
      \xxxSubParagraphStar
      \xxxSubParagraphNoStar
  }
  \newcommand{\xxxSubParagraphStar}[1]{\oldsubparagraph*{#1}\mbox{}}
  \newcommand{\xxxSubParagraphNoStar}[1]{\oldsubparagraph{#1}\mbox{}}
\fi
\makeatother

\usepackage{longtable,booktabs,array}
\usepackage{calc} 
\usepackage{etoolbox}
\makeatletter
\patchcmd\longtable{\par}{\if@noskipsec\mbox{}\fi\par}{}{}
\makeatother
\IfFileExists{footnotehyper.sty}{\usepackage{footnotehyper}}{\usepackage{footnote}}
\makesavenoteenv{longtable}
\usepackage{graphicx}
\graphicspath{{figures/}} 
\makeatletter
\def\maxwidth{\ifdim\Gin@nat@width>\linewidth\linewidth\else\Gin@nat@width\fi}
\def\maxheight{\ifdim\Gin@nat@height>\textheight\textheight\else\Gin@nat@height\fi}
\makeatother
\setkeys{Gin}{width=\maxwidth,height=\maxheight,keepaspectratio}
\makeatletter
\def\fps@figure{htbp}
\makeatother

\makeatletter
\@ifpackageloaded{caption}{}{\usepackage{caption}}
\AtBeginDocument{%
\ifdefined\contentsname
  \renewcommand*\contentsname{Table of contents}
\else
  \newcommand\contentsname{Table of contents}
\fi
\ifdefined\listfigurename
  \renewcommand*\listfigurename{List of Figures}
\else
  \newcommand\listfigurename{List of Figures}
\fi
\ifdefined\listtablename
  \renewcommand*\listtablename{List of Tables}
\else
  \newcommand\listtablename{List of Tables}
\fi
\ifdefined\figurename
  \renewcommand*\figurename{Figure}
\else
  \newcommand\figurename{Figure}
\fi
\ifdefined\tablename
  \renewcommand*\tablename{Table}
\else
  \newcommand\tablename{Table}
\fi
}
\@ifpackageloaded{float}{}{\usepackage{float}}
\floatstyle{ruled}
\@ifundefined{c@chapter}{\newfloat{codelisting}{h}{lop}}{\newfloat{codelisting}{h}{lop}[chapter]}
\floatname{codelisting}{Listing}

\makeatother
\makeatletter
\@ifpackageloaded{caption}{}{\usepackage{caption}}
\@ifpackageloaded{subcaption}{}{\usepackage{subcaption}}
\makeatother

\ifLuaTeX
  \usepackage{selnolig}  
\fi
\usepackage[]{natbib}
\usepackage{bookmark}

\IfFileExists{xurl.sty}{\usepackage{xurl}}{} 
\hypersetup{
  pdftitle={Title},
  pdfauthor={Author 1; Author 2},
  pdfkeywords={3 to 6 keywords, that do not appear in the title},
  colorlinks=true,
  linkcolor={blue},
  filecolor={Maroon},
  citecolor={Blue},
  urlcolor={Blue},
  pdfcreator={LaTeX via pandoc}}

\newcommand{\anon}{1}

\usepackage{empheq}

\usepackage{bm}
\usepackage{comment}
\usepackage{amsthm} 

\newcommand{\Cov}{\mathrm{Cov}}

\renewcommand{\th}{\mathrm{th}}

\newcommand{\Ltwo}{{L^2}} 
\newcommand{\normLtwo}[1]{\|#1\|_{L^2}} 
\newcommand{\innerLtwo}[2]{\langle #1,~#2 \rangle_{L^2}} 

\newcommand{\normH}[1]{\|#1\|_{\mathcal{H}}} 
\newcommand{\innerH}[2]{\langle #1, #2 \rangle_{\mathcal{H}}} 

\newtheorem{thm}{Theorem}
\newtheorem{prop}[thm]{Proposition}

\newtheorem{rmk}{Remark}

\newcommand{\rkhs}{\mathcal{H}}
\newcommand{\calM}{\mathcal{M}}

\renewcommand{\tilde}{\widetilde}

\newcommand{\blue}{\color{blue}}

\usepackage{xr}
\begin{document}

\def\spacingset#1{\renewcommand{\baselinestretch}%
{#1}\small\normalsize} \spacingset{1}


\if1\anon
{
  \title{\bf Blurring Mean Shift for Clustering Functional Data: A Scalable Algorithm and Convergence Analysis
  }
  \author{
  	Toshinari Morimoto$^{a}$,
  Ting-Li Chen$^{a}$\thanks{Corresponding author. Email: tlchen@stat.sinica.edu.tw},
  	Su-Yun Huang$^{a}$, and Ruey S. Tsay$^{b}$ \\
    $^a$Institute of Statistical Science, Academia Sinica, Taipei, Taiwan\\
  	$^b$Booth School of Business, University of Chicago, Chicago, IL, USA
  }
  \maketitle
} \fi

\if0\anon
{
  \bigskip
  \bigskip
  \bigskip
  \begin{center}
    {\LARGE\bf Mean Shift for Clustering Functional Data: A Scalable Algorithm and Convergence Analysis}
\end{center}
  \medskip
} \fi

\begin{abstract}
This paper extends the blurring mean shift algorithm from vector-valued data to functional data, enabling effective clustering in infinite-dimensional settings without requiring specification of the number of clusters.
To address the computational challenges posed by large-scale datasets, we introduce a fast stochastic variant that significantly reduces computational complexity.
We provide a rigorous convergence analysis for the full blurring functional mean shift procedure, establishing theoretical guarantees for its iterative behavior.
For the stochastic variant, we provide partial theoretical justification by showing that, when the subset size is sufficiently large, its one-step update is well approximated by the corresponding update of the full algorithm.
The proposed method is demonstrated through real-data applications, including hourly Taiwan PM$_{2.5}$ measurements and Argo oceanographic profiles.
Our key contributions include: 
(1) extending the blurring mean shift algorithm to functional data in a Hilbert-space setting;
(2) developing a scalable stochastic variant based on random partitioning for large-scale data; 
(3) establishing convergence results for the full blurring functional mean shift algorithm; and
(4) demonstrating the scalability and practical usefulness of the proposed method through simulation and real-data applications.
\end{abstract}
\noindent%
\vfill

\newpage
\spacingset{1.8} 

%
%

\section{Introduction}\label{sec:intro}

Mean shift is a nonparametric, mode-seeking algorithm for finding the modes of a density function \citep{fukunaga1975estimation} and has been widely applied in pattern recognition and image analysis \citep{cheng1995mean,Comaniciu2002}. 
Broadly speaking, the algorithm has two variants: the blurring and non-blurring versions.
In the blurring version, the data points that define the mean shift operator are updated at each iteration, whereas in the non-blurring version, the operator is defined with respect to the original data points throughout the iterative process.
The original mean shift algorithm of \citet{fukunaga1975estimation} is a blurring algorithm, whereas the version proposed by \citet{cheng1995mean} is a non-blurring algorithm.
Its ability to identify clusters with complex, non-convex shapes without requiring a prespecified number of clusters makes mean shift a powerful tool for clustering analysis. 
Unlike most traditional clustering methods that rely on a fixed number of clusters, mean shift directly seeks high-density regions and thus naturally adapts to complex data structures.
This flexibility and robustness have led to the broad adoption of the algorithm in computer vision, including applications such as image segmentation, object tracking, and other tasks requiring precise, adaptive clustering \citep{Comaniciu2002}.
In addition to its broad practical success, mean shift has also attracted growing theoretical interest in recent years.
In particular, \citet{yamasaki2024blurring} provided a rigorous analysis of the convergence behavior of the blurring mean shift algorithm in Euclidean spaces, further strengthening the theoretical foundation of the method.

In recent years, functional data have become increasingly common in a wide range of scientific fields, including biology, meteorology, economics, engineering, and medicine.
Such data often exhibit complex temporal or spatial patterns, together with high dimensionality and inherent smoothness. 
These features have motivated the development of new methods for functional data analysis.

While the mean shift algorithm has been extensively studied for vector-valued data, its functional counterpart has received comparatively less attention. 
Earlier works by \citet{ciollaro2014, ciollaro2016} laid important theoretical groundwork for the non-blurring functional mean shift algorithm.
They introduced a pseudo-density for functional data. 
They then studied a continuous-time gradient flow corresponding to the discrete mean shift update rule, under which each function moves toward a local mode of the pseudo-density over time.
Their main theoretical results established convergence of this {\it idealized continuous-time dynamics} under suitable regularity conditions.

In contrast, in this paper we study the blurring functional  mean shift algorithm for a finite sample of functional observations $f_1,\ldots,f_n$, with emphasis on the practical discrete update procedure applied to empirical data. 
Specifically, we consider the empirical analogue of the pseudo-density, obtained by replacing the underlying distribution with the empirical distribution of $f_1,\ldots,f_n$, and refer to it as the surrogate density. 
Rather than analyzing a continuous-time gradient flow, we study the discrete iterative updates of the blurring functional  mean shift algorithm itself. 
We show directly that the resulting iterative sequences converge, and that the limiting points are stationary and exhibit negative second-order curvature in the limit, thereby supporting their interpretation as local modes of the surrogate density. 
This provides a theoretical guarantee for the convergence and numerical stability of the actual algorithm implemented in practice.

A practical issue in applying mean shift to large datasets is its computational cost. 
Since each iteration of mean shift requires kernel-weighted averaging over the entire dataset, the computational complexity is of order $O(n^2)$, where $n$ is the number of observations.
To address this difficulty, we propose a fast stochastic variant of the functional mean shift algorithm, which substantially reduces the computational burden by operating on randomly partitioned subsets of the data at each iteration. 
This makes the method scalable to large-scale problems while preserving the qualitative behavior of the original algorithm.
The convergence guarantee described above is established for the full-data update rather than the randomized version, although we also provide a partial theoretical justification for the latter.

In summary, the main contributions of this work are:
\begin{enumerate}
\item
We formulate and study a blurring functional mean shift (BFMS) algorithm for finite samples of functional data in a Hilbert space setting, and establish convergence results for its discrete iterative update procedure.

\item
To handle large-scale functional datasets more efficiently, we propose a scalable stochastic BFMS algorithm based on random partitioning and cyclic subset-based updates.
This provides a practical alternative when full-data updates are computationally expensive.

\item
We provide a rigorous convergence analysis for the proposed BFMS algorithm in the full-data setting.
In contrast to earlier analyses of an {\it idealized continuous-time gradient flow}, our analysis directly concerns its practical discrete iterative updates.
This yields a direct theoretical guarantee for the convergence of the iterative procedure, as well as insight into the clustered structure of the limiting configuration ${f^{(\infty)}_1, \ldots, f^{(\infty)}_n}$, thereby supporting the practical use of BFMS for functional clustering.
We also provide a partial theoretical justification for the stochastic version by showing that, when the subset size is sufficiently large, its one-step subset-based updates approximate the corresponding full-data updates well.

\item 
We illustrate the practical usefulness of the proposed method through simulation and applications to hourly Taiwan PM$_{2.5}$ measurements and Argo oceanographic profiles.
These examples show that the method can produce interpretable clusters in real functional data, with the Argo application also illustrating its scalability to larger datasets.
\end{enumerate}

Although both the full-data functional mean shift algorithm and the fast stochastic variant are formulated in general Hilbert spaces, our convergence analysis is specifically developed for the $\Ltwo([0,1])$ setting. 
Nevertheless, these results may extend to other separable Hilbert spaces under suitable conditions.

The rest of the article is organized as follows. 
Section~\ref{sec:methodology} introduces the functional mean shift algorithm along with its fast stochastic variant. 
Section~\ref{sec:convergence} presents the convergence analysis and Section~\ref{sec:simulation} provides simulation results, demonstrating convergence and scalability.
Section~\ref{sec:airbox} and Section~\ref{sec:argo} demonstrate the applications of our method to Taiwan hourly PM$_{2.5}$ measurements and Argo data clustering, respectively. 
The Argo profiles have more than one million functions of temperature and salinity across all oceans, presenting a large-scale computational challenge.
Finally, Section~\ref{sec:conclusion} concludes the article with a brief discussion.
All technical details are deferred to the Supplementary Material.
Part~A reviews the G\^ateaux derivatives used in the convergence analysis, and Part~B contains the technical proofs.

%
%

\section{Methodology}\label{sec:methodology}

The functional mean shift algorithm extends the traditional mean shift approach \citep{cheng1995mean} to handle functional data, enabling mode-seeking in infinite-dimensional spaces. 
To briefly recall, the traditional algorithm starts with a set of observed points $\{\bm{x}_i\}_{i=1}^n$ in $\mathbb{R}^p$, and constructs a kernel density estimate of the underlying distribution.  
Each point is iteratively updated toward a mode, that is, a local maximum of the estimated density, leading the points to gradually concentrate in high-density regions and ultimately form clusters.  
In contrast, functional data lie in infinite-dimensional spaces, where the probability density function (pdf) cannot be defined.  
To address this difficulty, we introduce the notion of a ``\emph{surrogate density}'' \citep{Ferraty2012}, which plays a similar role as the empirical pdf in guiding the mode-seeking process.  
Using this surrogate, we  extend the mean shift framework to the functional setting.

\subsection{Functional mean shift operator on a Hilbert space} 

\subsubsection{Definition of the functional mean shift operator}

Let $\{f_i\}_{i=1}^n$ be a set of functional observations in a Hilbert space $\rkhs$ equipped with an inner product $\innerH{\cdot}{\cdot}$.  
Assume that each function $f_i$ is defined over a common domain ${\cal T}$.
Given $\{f_i\}_{i=1}^n$, we introduce a surrogate density function $\rho(\cdot\mid\{f_i\}_{i=1}^n): \rkhs \to [0,\infty)$ below to serve as a proxy for the notion of density in infinite-dimensional spaces:
\begin{equation}\label{eq:surrogate}
\rho(f |\{f_i\}_{i=1}^n)=\frac1{n} \sum_{i=1}^n K_h(\normH{f - f_i}),
\end{equation}
where $\normH{\cdot} = \innerH{\cdot}{\cdot}^{1/2}$ denotes the norm on $\rkhs$ induced by the inner product and $\normH{f - f_i}$ therefore represents the distance between $f$ and $f_i$. 
The function $K_h(\cdot) = h^{-1} K(\cdot / h)$ is a univariate Gaussian kernel.
Other kernel functions can be used, but for simplicity, we adopt the Gaussian kernel.
This formulation can be viewed as a natural extension of kernel density estimation to infinite-dimensional settings.

Based on the surrogate density introduced in~\eqref{eq:surrogate}, the functional mean shift (FMS) operator is defined for functions in $\rkhs$.  
Given $\{f_i\}_{i=1}^n \subset \rkhs$, the mean shift operator $\calM(\cdot \mid \{f_i\}_{i=1}^n): \rkhs \to \rkhs$ is defined as follows:
\begin{equation}\label{eq:MS_operator}
\mathcal{M}(f \mid \{f_i\}_{i=1}^n) = \frac{\sum_{i=1}^n K_h\left(\normH{f - f_i}\right) f_i}{\sum_{i=1}^n K_h\left(\normH{f - f_i}\right)}.
\end{equation}
This operator plays the role of a mode-seeking mechanism in $\rkhs$, serving as an infinite-dimensional counterpart of the mean shift update rule in the traditional algorithm.
In particular, applying $\mathcal{M}(\cdot\mid\{f_i\}_{i=1}^n)$ to each observed function $f_i$ yields a new function $f^{(\rm new)}_i$ that is shifted toward a mode, i.e., a local maximum, of the surrogate density $\rho(\cdot \mid \{f_i\}_{i=1}^n)$ near $f_i$.  
As will be explained in Section~\ref{sec:two_variants_of_FMS}, this operation is repeatedly applied to all $f_1, \ldots, f_n$.  
Through this iterative process, each function $f_i$ gradually moves and eventually settles at a fixed point. 
Functions that converge to the same limit point are grouped into the same cluster, and clustering is thereby achieved.  
The derivation of the FMS operator is provided in the online Supplementary Material.

\subsubsection{Two variants of iterative updates with FMS operator}\label{sec:two_variants_of_FMS} 

The functional mean shift operator introduced in~\eqref{eq:MS_operator} is applied to each function in $\{f_i\}_{i=1}^n$ at every iteration, producing updated functions $\{f^{(\nu)}_i\}_{i=1}^n$, where $\nu$ denotes the current iteration number.  
For each fixed $i = 1, \ldots, n$, the sequence $\{f^{(\nu)}_i\}_{\nu=1}^\infty$ is expected to eventually converge to a limit point $f^{(\infty)}_i$.  
Functions that converge to the same point, e.g., $f_{i_1}, \ldots, f_{i_k}$ such that $f^{(\infty)}_{i_1} = \ldots = f^{(\infty)}_{i_k}$, are considered to belong to the same cluster, and clustering is then achieved 
(in practice, the iteration is terminated once the updates become stable).  

There are two variants of this iterative scheme, which differ in whether the surrogate density is consistently defined using the original functions $\{f_i\}_{i=1}^n$, or updated at each iteration based on the current functions $\{f_i^{(\nu)}\}_{i=1}^n$.
\begin{itemize}
\item
(\textbf{NBFMS}) The first variant is referred to as the non-blurring functional mean shift (abbreviated as NBFMS, or simply FMS):
\begin{equation}\label{eq:NBfMS}
	f^{[\nu +1]} = \calM (f^{[\nu]} \mid \{f_i\}_{i=1}^n),
\end{equation}
where $\nu$ denotes the current iteration number.  
In NBFMS, the underlying surrogate density is consistently defined using the original functions $\{f_i\}_{i=1}^n$.  

\item
(\textbf{BFMS}) The second variant is the blurring type (abbreviated as BFMS):
\begin{equation}\label{eq:BfMS}
	f^{(\nu +1)} = \calM (f^{(\nu)} \mid \{f_i^{(\nu)}\}_{i=1}^n).
\end{equation}
In contrast, in BFMS the underlying surrogate density is redefined at each iteration step based on the current functions $\{f_i^{(\nu)}\}_{i=1}^n$.
\end{itemize}

Note that, in Equations~\eqref{eq:NBfMS} and~\eqref{eq:BfMS}, $f^{(\nu)}$ (or $f^{[\nu]}$) on the right hand side should be understood as representing one of the current version of functions $f_i^{(\nu)}$ (or $f_i^{[\nu]}$) in the iterative process.
At the initial step, each $f_i^{(0)}$ (or $f_i^{[0]}$) is set to the original function $f_i$.
By repeatedly applying these update rules (i.e., $\nu=1,2,\ldots$), then each $f^{(\nu)}_i$ (or $f^{[\nu]}_i$) will eventually converge to a fixed point $f^{(\infty)}_i$ (or $f^{[\infty]}_i$).

The choice between NBFMS and BFMS depends on the specific application and desired properties of the clustering process. 
In this article we focus on the BFMS approach due to its faster convergence compared to NBFMS.

The above update formulas can be directly applied to small-scale functional datasets.  
However, since each iteration requires computing pairwise distances among the $n$ functions, the computational cost per iteration is $O(n^2)$.  
While this is manageable for small $n$, it becomes prohibitive for large-scale data.  
To address this issue, we develop an efficient computational strategy in Section~\ref{fast_alg}.

\subsection{Stochastic fast algorithm for the BFMS operator}\label{fast_alg}

To improve computational efficiency of the BFMS algorithm for large-scale functional data, we adopt a stochastic algorithm in which a fresh random partition of the data is generated at each iteration, following the approach in \citet{shiu2024randomized}. 
In this scheme, each functional observation, i.e., a point in the underlying Hilbert space $\rkhs$, interacts only with the points within its assigned random subset. 
By limiting pairwise distance computations to within subsets, this strategy substantially reduces the computational burden at each iteration.

\subsubsection{Description of the algorithm}

Given a large dataset $\{f_i\}_{i=1}^n$, the randomized algorithm is designed to efficiently approximate the BFMS operator while maintaining accuracy in clustering. The algorithm starts with the original dataset as initial state: $f_i^{(0)} =f_i$, for $i=1,\dots,n$.

\begin{itemize}
\item
{\bf Step 1: Random partitioning.} 
At the $\nu^\text{th}$ iteration, the full dataset is randomly partitioned into $m$ disjoint subsets, denoted by $D^{(\nu)}=\cup_{k=1}^m D^{(\nu)}_{{\cal J}_k}$, each containing approximately the same number of elements. 
Then, to update each blurred data point, $f_i^{(\nu)}$, instead of utilizing the entire dataset, we only use the subset $D^{(\nu)}_{{\cal J}_{k(i)}}$, to which $f_i^{(\nu)}$ belongs, i.e., $f_i^{(\nu)}\in D^{(\nu)}_{{\cal J}_{k(i)}}$. 
This random partitioning strategy allows us to approximate the surrogate density using a smaller, computationally manageable portion of the data while maintaining representativeness (refer to~\eqref{surrogate_approx} below).

\item
{\bf Step 2: Mean shift operator based on a stochastic subset.} 
More specifically, to update the point $f_i^{(\nu)}$, which belongs to the random subset $D^{(\nu)}_{{\cal J}_{k(i)}}$, the full data BFMS operator is approximated using only this subset. 
The BFMS approximation is given by:
\begin{equation}\label{eq:rBfMS}
	\mathcal{M}(f_i^{(\nu)} | D^{(\nu)}) \approx\calM (f_i^{(\nu)}  | D^{(\nu)}_{{\cal J}_{k(i)}})  
	= \frac{\sum_{j\in {\cal J}_{k(i)}} K_h(\normH{f_i^{(\nu)}  - f_j^{(\nu)}})~f_j^{(\nu)}}
	{\sum_{j\in {\cal J}_{k(i)} }  K_h(\normH{f_i^{(\nu)}-f_j^{(\nu)}})}, 
\end{equation}
where ${\cal J}_{k(i)}$ denotes the index set of the subset containing $f_i^{(\nu)}$. 
(A more accurate notation should be ${\cal J}_{k(i)}^{(\nu)}$, but for simplicity, we use ${\cal J}_{k(i)}$.)
This approximation significantly reduces computational cost while maintaining an accurate surrogate density estimate, provided that each subset contains sufficient functional data.

\item
{\bf Step 3: Iterative update with stochastic BFMS.}
Starting with the original dataset as the initial state, i.e., $f_i^{(0)} = f_i$ for $i = 1, \dots, n$, each point is updated iteratively using the stochastic BFMS operator defined in~\eqref{eq:rBfMS}:
\[ f_i^{(\nu+1)} = \calM (f_i^{(\nu)}  | D^{(\nu)}_{{\cal J}_{k(i)}}), \]
where ${\cal J}_{k(i)}$ denotes the index set of the subset $D_{{\cal J}_{k(i)}}^{(\nu)}$ that contains the current point $f_i^{(\nu)}$.
The kernel value $K_h(d(f_i^{(\nu)}, f_j^{(\nu)}))$, used in $\calM (f_i^{(\nu)}  | D^{(\nu)}_{{\cal J}_{k(i)}})$, quantifies the similarity between the current point $f_i^{(\nu)}$ and each point $f_j^{(\nu)}$ in the random subset $D^{(\nu)}_{{\cal J}_{k(i)}}$, to which $f_i^{(\nu)}$ belongs. 
This weighted average shifts $f_i^{(\nu)}$ towards regions of higher density, as represented by $D^{(\nu)}_{{\cal J}_{k(i)}}$. 
The corresponding surrogate density evaluated at $f_i^{(\nu)}$ can be approximated by:
\begin{eqnarray}
	&& \rho\left(f_i^{(\nu)} \big| D^{(\nu)}\right)
	= \frac1{n} \sum_{j=1}^n K_h\left(\normH{f_i^{(\nu)} - f_j^{(\nu)}}\right)\nonumber\\
	&\approx& \rho\left(f_i^{(\nu)} \big| D^{(\nu)}_{{\cal J}_{k(i)}}\right)
	= \frac1{n_i} \sum_{j\in {\cal J}_{k(i)}} K_h\left(\normH{f_i^{(\nu)} - f_j^{(\nu)}}\right) ,
	\label{surrogate_approx}
\end{eqnarray}
where $n_i$  is the size of the subset $D^{(\nu)}_{{\cal J}_{k(i)}}$ (i.e., the cardinality of ${\cal J}_{k(i)}$). 
This stochastic formulation provides an efficient approximation to both the full-data surrogate density estimate and the mean shift operator, substantially reducing computational complexity. 
The iterative procedure continues until the updates stabilize, that is, until the update magnitude falls below a predefined threshold $\epsilon$:
\[
\normH{f^{(\nu+1)} -  f^{(\nu)}}< \epsilon.
\]
This stopping criterion ensures that the current estimate $f^{(\nu)}$ has reached a stationary point in the function space $\rkhs$, where the mean shift dynamics have converged.

\item
{\bf Step 4: Cluster membership assignment.} 
After the iterative updates, each function $f_i$ eventually converges to a limiting point $f_i^{(\infty)}$.
Functions that share the same limiting point, i.e., $f_{i_1}, \ldots, f_{i_k}$ such that $f_{i_1}^{(\infty)} = \cdots = f_{i_k}^{(\infty)}$, are assigned to the same cluster.
\end{itemize}

Our stochastic BFMS algorithm differs fundamentally from standard mini-batch optimization methods such as mini-batch stochastic gradient descent (SGD). 
In conventional mini-batch approaches, small subsets of data are sampled at each iteration, and updates are performed one mini-batch at a time, typically in a sequential or epoch-based manner that cycles through all mini-batches. 
In contrast, our algorithm partitions the full dataset into disjoint subsets (which may be viewed as mini-batches) at each iteration, and each subset is used to simultaneously update the mean shift estimates for the functional data points it contains. 
As a result, all subsets contribute updates in parallel within the same iteration, and the entire dataset is processed without sequential cycling. 
This design enables efficient, distributed computation while preserving the core mode-seeking behavior of the functional mean shift algorithm.

\subsubsection{Computational complexity}\label{complexity}

By using a subset of size $\tilde{n} \approx n/m$ in place of the full dataset, the computational complexity of the mean shift update is reduced from $O(n^2)$ to $O(\tilde{n} n)$. Since $\tilde{n} \ll n$, this results in substantial computational savings, making the proposed stochastic algorithm efficient and practical for clustering large-scale functional datasets.

\subsection{Handling partially observed trajectories}

In practice, each trajectory $f_i(t)$ is typically observed only at a subset of discrete time points, resulting in partially observed functional data. 
While this poses additional challenges for clustering, the primary focus of this paper is on developing a fast stochastic algorithm for large-scale functional clustering. 
To stay focused on this main objective, we leave the treatment of partially observed trajectories for future work, while briefly addressing them in the Argo data analysis as part of the data preprocessing in the real-data application.

%
%

\section{Convergence analysis}\label{sec:convergence}

In this section, we present theoretical results on the convergence of the BFMS algorithm under the full-data update procedure, together with the resulting clustered structure as characterized by the separation between cluster centers.
We also provide a partial theoretical justification for the stochastic variant.

\subsection{Main convergence theorem}

In the following discussion, for simplicity, we consider $\mathcal{T} = [0,1]$ and assume $\rkhs = \Ltwo([0,1])$, equipped with the inner product $\innerLtwo{f}{g} = \int_0^1 f(t) g(t)\,dt$, which induces the norm $\normLtwo{f}^2 = \innerLtwo{f}{f}$.
Furthermore, we impose the following condition on the kernel function:
\begin{equation}\label{condition}
    \text{$K_h$ is a univariate kernel with bandwidth $h$ and compact support $[-\tau, \tau]$.}
\end{equation}

For simplicity, we adopt the truncated Gaussian kernel:
\begin{equation}\label{eq:truncated_Gaussian}
K_h(t) = \frac{1}{\sqrt{2\pi}\, h} e^{- t^2 / 2h^2} \, \mathcal{I}(|t| \leq \tau),
\end{equation}
where ${\cal I}$ is the indicator function.
With these assumptions, we now establish the following key properties of BFMS.

\begin{thm}[Convergence properties]\label{thm:main}
Assume $\rkhs=\Ltwo([0,1])$ and that condition~\eqref{condition} holds. Then,
\renewcommand{\labelenumi}{(\Alph{enumi})}
\begin{enumerate}
\item 
{\bf Monotonic increase of the average surrogate density.} The average surrogate density $\rho(F^{(\nu)})$ given by  
\begin{equation}\label{average_density}
	\rho(F^{(\nu)}) = \frac{1}{n} \sum_{j=1}^n \rho(f_j^{(\nu)} \mid F^{(\nu)}) 
\end{equation}
is monotonically increasing in the iteration index $\nu$, where $F^{(\nu)}=\{f_i^{(\nu)}\}_{i=1}^n$ denotes the updated configuration in $\Ltwo([0,1])$ at $\nu^\th$ iterations. 

\item
{\bf Convergence of the BFMS process.} 
For each $i=1,\ldots,n$, the sequence $\{f_i^{(\nu)}\}_{\nu=1}^\infty$ converges  in $\Ltwo$, i.e., there exists $f^{(\infty)}_i \in \Ltwo([0,1])$ such that $\normLtwo{f^{(\nu)}_i - f^{(\infty)}_i} \to 0$ as $\nu \to \infty$.

\item 
{\bf Limiting points are stationary and correspond to modes.}
Let $f_i^{(\infty)}$ denote the limit of the sequence $\{f_i^{(\nu)}\}_{\nu=1}^\infty$, and define the limiting configuration by
\[
F^{(\infty)} = \{f_1^{(\infty)}, f_2^{(\infty)},\ldots,f_n^{(\infty)}\}.
\]
Then, each limiting point $f_i^{(\infty)}$ is a stationary point in the following limiting sense:
\begin{itemize}
\item First,
\begin{equation}\label{limit_stationary}
\lim_{\nu\to\infty} \delta \rho(f \mid F^{(\nu)}) [g]\big|_{f = f_i^{(\nu)}} = 0, \qquad \forall i,\ \forall g\in L^2([0,1]),
\end{equation}
where $\delta \rho(f \mid F^{(\nu)})[g]$ denotes the first-order G\^ateaux derivative of $\rho$ with respect to $f$ in the functional direction $g$.
\item
Moreover, the second-order G\^ateaux derivative of $\rho(f \mid F^{(\nu)})$ with respect to $f$, evaluated along arbitrary functional directions $(g_1,g_2)$, defines a bilinear form. In particular, for every nonzero direction $g$,
\[
\lim_{\nu\to\infty} \delta^2 \rho(f \mid F^{(\nu)})[g,g]\big|_{f=f_i^{(\nu)}} < 0, \quad \forall i,
\]
so the second-order G\^ateaux derivative is strictly negative definite in the limit.
\end{itemize}
\item
\textbf{Structure of limiting points:} 
The limiting configuration $F^{(\infty)}$ consists of $k$ distinct cluster centers $(k \leq n)$, denoted by $\{v_c\}_{c=1}^k$, where $v_1,\dots,v_k \in L^2([0,1])$.
These centers are mutually separated by at least $\tau$, in the sense that $\normLtwo{v_c - v_{c'}} \ge \tau$ for all $c \neq c'$. 
\end{enumerate}
\end{thm} 



Theorem~\ref{thm:main} shows that, under the full-data BFMS update, the average surrogate density over the evolving configuration is nondecreasing across iterations. 
It also establishes that each sequence $f_i^{(\nu)}$ converges as $\nu\to\infty$, which provides a natural stopping criterion in practice, since $\normLtwo{f_i^{(\nu+1)}-f_i^{(\nu)}} \to 0$. 
Moreover, although the surrogate density itself evolves with the iteration, the limiting points are stationary and exhibit negative second-order curvature in the limit, so they may be regarded as local modes. 
Finally, the limiting configuration consists of a finite collection of distinct and well-separated cluster centers in $L^2([0,1])$, with functions converging to the same limit assigned to the same cluster.

We conclude this section with a remark on the scope of applicability of the proposed method.
\begin{rmk}[applicability to a broad class of functional data]
The functional mean shift operator and the convergence analysis developed here require only square integrability, namely, $\normLtwo{f}<\infty$, and therefore do not rely on additional smoothness assumptions such as the existence of $f'$.
Accordingly, the proposed method applies to a broad class of functional data.
This is particularly relevant in applications, where functions are often observed with noise or only partially, so that strong pointwise smoothness assumptions may be difficult to justify.
However, if one wishes to carry out post-clustering pointwise inference, such as evaluating function values at specific time points, then additional regularity conditions are required. For instance, one may assume that the functions belong to a reproducing kernel Hilbert space, in which pointwise evaluation is a continuous linear functional.
\end{rmk}


\subsection{Justification for the stochastic variant}

In Theorem~\ref{thm:main}, we established the foundational convergence properties of the full-data BFMS algorithm. 
A natural question that arises is whether the stochastic update sequence converges to the full-data update sequence as $\tilde{n}= n/m \to \infty$. 
While a complete convergence theory for the stochastic iterates is technically challenging and remains unresolved in the current work, the following proposition provides partial theoretical justification. 
It shows that, when the subset size is sufficiently large, the one-step stochastic BFMS update provides a valid approximation to the one-step full-data update.

\begin{prop}[LLN for the subset-based FMS operator]\label{prop:general_fLLN}
Let $\{g_j\}_{j=1}^\infty \subset \Ltwo([0,1])$ be a sequence of functions satisfying $\normLtwo{g_j} < C$ for some constant $C$. 
Suppose the kernel $K_h$ satisfies condition~\eqref{condition}.
Define the full-data mean shift operator based on $\{g_j\}_{j=1}^n$ as
\[
\calM(f \mid \{g_j\}_{j=1}^n) = \frac{\sum_{j=1}^n K_h(\normLtwo{f - g_j}) g_j}{\sum_{j=1}^n K_h(\normLtwo{f - g_j})}.
\]
Let $\mathcal{J} \subset \{1, \dots, n\}$ be a uniformly drawn subset of size $\tilde{n} = n/m$, where $m$ is the number of partitions.
Without loss of generality, we assume that $\tilde{n}$ is an integer. 
Suppose $m = m(n)$ satisfies $\tilde{n} \to \infty$ as $n \to \infty$. 
Define the subset-based (partial-data) mean shift operator as
\[
\calM(f \mid \{g_j\}_{j\in \mathcal{J}}) = \frac{\sum_{j \in \mathcal{J}} K_h(\normLtwo{f - g_j}) g_j}{\sum_{j \in \mathcal{J}} K_h(\normLtwo{f - g_j})}.
\]
Then, we have
\begin{equation}\label{eq:fLLN}
	\normLtwo{
		\calM(f \mid \{g_j\}_{j\in \mathcal{J}}) - 
		\calM(f \mid \{g_j\}_{j=1}^n) 
	} 
	\to 0
	\quad \text{in probability as } \tilde{n} \to \infty.
\end{equation}
\end{prop}

%
%

\section{Simulation Study}\label{sec:simulation}

In this section, we examine the efficacy of our clustering method in finite functional samples via simulations.
We generate functional data from distinct mean functions corresponding to different clusters and apply the proposed algorithm to examine whether it can successfully recover the underlying cluster structure.


\subsection{Simulation settings}\label{sec:simulation_settings}

In this section, we describe how the functional data are generated and specify the parameter settings used in the simulations.

\subsubsection{Data generating process}

We generate functional data by adding noise to distinct mean functions corresponding to different clusters.
Each function is represented on an equally spaced grid of evaluation points, as described below.

\medskip
\noindent\textbf{1. Domain and discretization.}\\
All functions are defined on the interval $\mathcal{T} = [0,1]$ and observed on an equally spaced grid of $p+1$ points, which partition $\mathcal{T}$ into $p$ subintervals.
In all simulation settings, we fix $p = 200$.

\medskip
\noindent\textbf{2. Cluster structure.}\\
In this simulation study, we consider $K=4$ clusters.
Each cluster is associated with a distinct mean function $\mu_k(t)$, where $k=1, \ldots, K$.
For each cluster, $n_{\rm per} = 5{,}000$ curves are generated by adding random noise components (described below) to the corresponding mean function, resulting in a total of $n = 20{,}000$ functional observations.  
The cluster membership of the $i^\th$ curve is denoted by $z_i \in \{1,\ldots, K\}$.

\medskip
\noindent\textbf{3. Function generation model.}\\
Each observation $f_i(t)$ is generated according to
\[
f_i(t) = \mu_{z_i}(t) + \eta_i(t) + \epsilon_i(t),
\]
where $\eta_i(t)$ is a smooth random fluctuation and $\epsilon_i(t)$ is Gaussian white noise with mean zero and variance $\sigma_{\mathrm{white}}^2$, generated independently at each discretized time point.  
To define the mean functions, we introduce two template functions: a Gaussian bump centered at $c$ with width $w$ and a sigmoid function centered at $c$ with slope parameter $a$, given by
\[
\phi_{\mathrm{G}}(t; c, w) = \exp\!\left( -\frac{(t-c)^2}{2w^2} \right), \quad
\phi_{\mathrm{S}}(t; c, a) = \frac{1}{1 + \exp\{-a(t - c)\}}.
\]
The four cluster mean functions differ in peak location, modality, and periodicity, are given by
\begin{align*}
\mu_1(t) &= \phi_{\mathrm{G}}(t; 0.50, 0.05),\\
\mu_2(t) &= 0.5\,\phi_{\mathrm{G}}(t; 0.30, 0.05)
+ 0.5\,\phi_{\mathrm{G}}(t; 0.70, 0.05),\\
\mu_3(t) &= 2\phi_{\mathrm{S}}(t; 0.50, 16)-1,\\
\mu_4(t) &= \cos(4\pi t).
\end{align*}
To add smooth random fluctuations around the mean function, we generate each $\eta_i$ from a Gaussian process. Specifically, for each $i$, the process $\{\eta_i(t): t\in\mathcal{T}\}$ satisfies
\[
\mathbb{E}[\eta_i(t)]=0,\quad \Cov\left(\eta_i(t),\eta_i(t')\right)=k_{\mathrm{Mat\acute ern}}(t,t'),
\]
where
\[
k_{\mathrm{Mat\acute ern}}(t,t')
=\sigma_{\mathrm{smooth}}^2\,
\frac{2^{1-\nu}}{\Gamma(\nu)}
\!\left(\frac{\sqrt{2\nu}\,|t-t'|}{\ell}\right)^{\!\nu}
K_{\nu}\!\left(\frac{\sqrt{2\nu}\,|t-t'|}{\ell}\right).
\]
Here, $K_\nu$ denotes the modified Bessel function of the second kind, $\ell$ is the length-scale, and $\nu$ controls smoothness. On an equally spaced grid $t_0,\ldots,t_p$, the discretized sample paths are generated as
\[
\bm{\eta}_i=\big(\eta_i(t_0),\ldots,\eta_i(t_p)\big)^\top \sim \mathcal{N}\left(\mathbf{0},\mathbf{K}\right),
\quad \mathbf{K}_{ij}=k_{\mathrm{Mat\acute ern}}(t_i,t_j),
\]
where $\sigma_{\mathrm{smooth}}=0.1$, $\ell=1.0$, and $\nu=2.5$.
In addition to this smooth component, we further add a noise term to represent local measurement error.
Specifically, the white noise term $\epsilon_i(t)$ is generated independently at each grid point from $\mathcal{N}(0, \sigma_{\mathrm{white}}^2)$ with $\sigma_{\mathrm{white}} = 0.1$.

\begin{figure}[H] 
\centering
\includegraphics[width=0.75\textwidth]{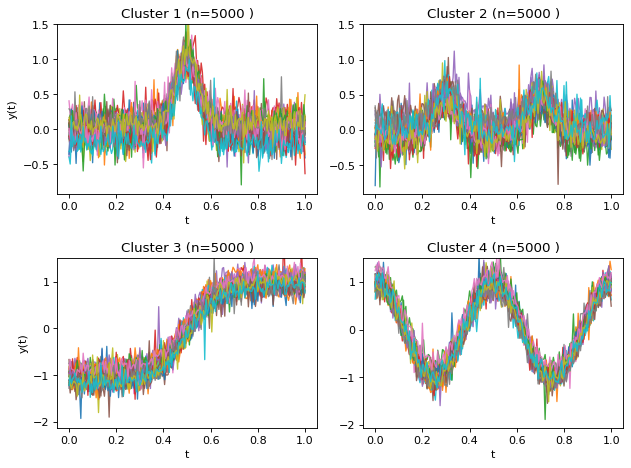}
\caption{Clustered functional data generated in the simulation study. 30 randomly selected curves are shown for each cluster.}
\label{fig:simulation_clusters}
\end{figure}

\medskip
\noindent\textbf{4. Illustration of generated functional data.}\\
Figure~\ref{fig:simulation_clusters} presents examples of the functional observations generated in the simulation study, with up to 30 sample curves per cluster.
The plots clearly reflect the underlying cluster mean functions while also exhibiting the smooth Gaussian-process fluctuations and added white noise.

\subsubsection{Clustering settings}

Instead of directly performing clustering, we first smooth each curve using a moving average with window size 5, keeping its length unchanged by padding the endpoints through edge-value extension.  
We then apply the stochastic BFMS algorithm described in Section~\ref{fast_alg} with the truncated Gaussian kernel defined in~\eqref{eq:truncated_Gaussian}, using the following settings:
\begin{itemize}
\item 
The bandwidth schedule is given by \eqref{eq:bandwith_schedule}, where $\nu$ denotes the iteration number.
Note that $h_\nu$ increases with $\nu$ during the first 20 iterations and remains fixed thereafter.
Thus, in the early stages, a smaller bandwidth allows finer-scale data movement and helps preserve smaller-scale cluster structure, whereas in the later stages, a larger bandwidth promotes the merging of nearby groups into broader and more stable clusters.
This schedule is adopted throughout the subsequent sections.
\begin{equation}\label{eq:bandwith_schedule}
h_\nu=\frac{\tau}{100}\bigl(5+2\min(\nu,20)\bigr).
\end{equation}

\item
Figure~\ref{fig:simulation_histogram_distance} shows the histogram of pairwise distances between 1,000 randomly selected curves in one repeated run.
The first peak mainly corresponds to within-cluster distances, while the second peak arises from distances between neighboring cluster centers.
Here, $\tau$ serves as the radius of the influential range of the kernel: within-cluster distances are typically smaller than $\tau$, whereas inter-cluster distances tend to exceed $\tau$.
Therefore, we set $\tau$ to the valley between the first and second peaks, which leads us to adopt $\tau = 3.5$.

\begin{figure}[H]
	\centering
	\includegraphics[width=0.55\textwidth]{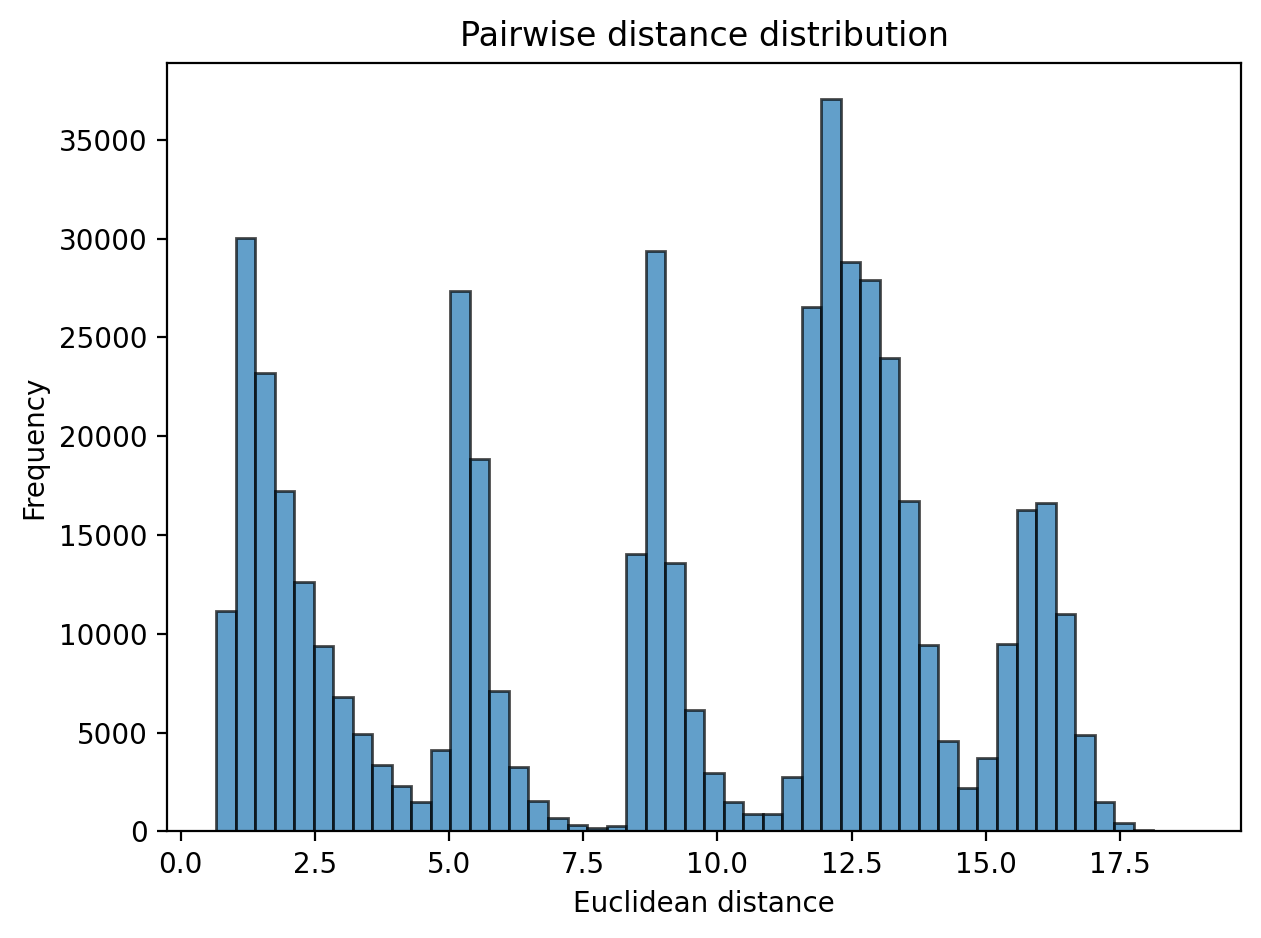}
	\caption{Histogram of pairwise distances between 1,000 randomly sampled functions.}
	\label{fig:simulation_histogram_distance} 
\end{figure}

\item 
For the stochastic variant, the data are randomly split at each iteration into $m = \lceil n / \tilde{n} \rceil$ disjoint groups.
We consider $\tilde{n} \in \{250, 500, 1000, 2000\}$ in our simulations in order to examine the effect of the subset size on both clustering accuracy and computation time.
Smaller values of $\tilde{n}$ correspond to more partitions, which are expected to reduce computation times, while larger values may improve robustness of the results.
\end{itemize}


\subsection{Simulation results}

\subsubsection{Organization of the simulation results figure}

\begin{figure}[H]
\centering
\includegraphics[width=\textwidth]{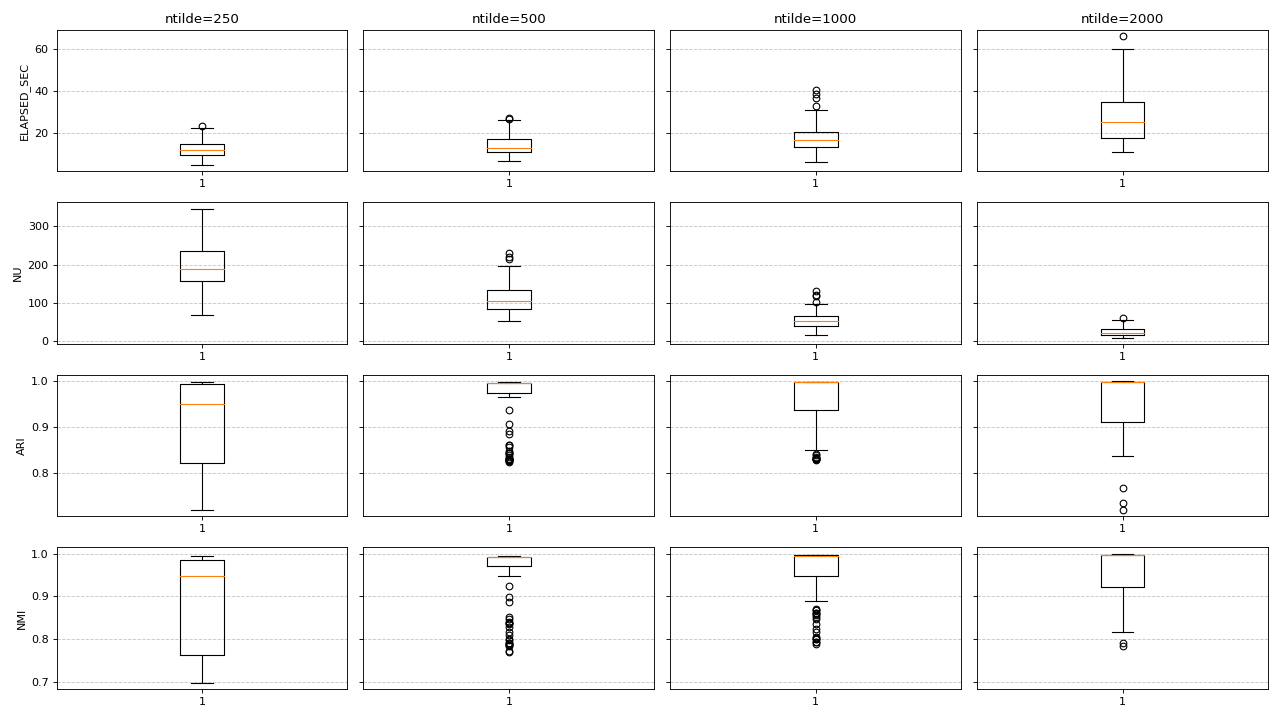}
\caption{
Boxplots of computation time (top row), the number of iterations $\nu$ until convergence (second row), Adjusted Rand Index (third row), and Normalized Mutual Information (bottom row) for $\tilde{n} \in \{250, 500, 1000, 2000\}$ over 100 trials.
}
\label{fig:simulation_boxplot}
\end{figure}

Figure~\ref{fig:simulation_boxplot} summarizes the results of 100 independent trials for each subset size $\tilde{n} \in \{250, 500, 1000, 2000\}$.
Since both the data generation process and the random partitioning in the stochastic fast algorithm vary across trials, each panel is presented as a boxplot to summarize the variability across the 100 trials.
The contents of the figure are organized as follows.
\begin{enumerate}
\item The top row shows computation time in seconds required to complete the clustering.
\item The second row shows the number of iterations $\nu$ required until the clustering procedure stops.
\item The third row shows the Adjusted Rand Index (ARI) \citep{Hubert1985}, which evaluates the similarity between the true and estimated clusters by checking whether pairs of items are consistently assigned to the same or different clusters, with a correction for chance agreement.
\item The fourth row shows the Normalized Mutual Information (NMI), which treats the estimated and true clusterings as two distributions, measures their divergence via the KL distance (mutual information), and normalizes it by their entropies.
\end{enumerate}
Both ARI and NMI are invariant to label permutations, and higher values indicate better agreement between the estimated and true clusterings.

\subsubsection{Observations and possible explanations}

\noindent\textbf{1. Computational time.}

The elapsed computation time required to complete clustering is similar for $\tilde{n}=250$ and $500$, but tends to increase for the larger values $\tilde{n} = 1000$ and $2000$.
By contrast, the number of iterations $\nu$ required until the algorithm stops decreases as $\tilde{n}$ increases.
This may be because each update is computed from a larger subset, which makes the update more stable and thus accelerates convergence.
At the same time, a larger $\tilde{n}$ increases the computational cost of each update.
As a result, although fewer iterations are required, the overall elapsed time tends to increase when $\tilde{n}$ is set to relatively large values such as $1000$ or $2000$.

\noindent\textbf{2. Clustering performance.}

For $\tilde{n}=250$, the median values of both ARI and NMI are already around 0.95, indicating that the stochastic fast algorithm achieves reasonably accurate clustering even with a relatively small subset size.
However, the variability of these scores across trials is larger, especially on the lower side.

For $\tilde{n}=500,1000$, and $2000$, the median values of both ARI and NMI are nearly equal to 1.
However, the stability of these scores does not improve monotonically beyond $\tilde{n}=500$.
In particular, the interquartile ranges for $\tilde{n}=500$ appear smaller than those for $\tilde{n}=1000$ or $2000$.
One possible explanation is that, as $\tilde{n}$ increases, the number of partitions decreases, so the effect of random partitioning becomes more pronounced, which may lead to greater variability in the clustering result.

\noindent\textbf{3. Practical implications for large-scale data.}

These observations suggest that, for sufficiently large $n$, the subset size $\tilde{n}$ should be kept as small as possible while still ensuring stable subset-based updates.
This choice keeps the number of partitions sufficiently large, which helps suppress the randomness induced by random partitioning.
It therefore provides a favorable balance between computational efficiency and clustering stability.

%
%

\section{Application to AirBox PM${_{2.5}}$ data}\label{sec:airbox} 

In Sections~\ref{sec:airbox} and~\ref{sec:argo}, we apply the proposed method to real-world datasets. 
In this section, we begin with the Taiwan AirBox dataset, which, though relatively small and not intended to test scalability, serves as a convenient initial testbed due to its manageable size and our familiarity with the local geography.

%
%

\subsection{Dataset, preprocessing, and clustering configuration}

\subsubsection{Dataset}

We use the Taiwan AirBox dataset \citep{Chen2017_PM25}, which records hourly PM$_{2.5}$ concentrations at 516 monitoring sites across Taiwan during March 2017. 
The dataset comprises 744 time points, corresponding to the 31 days of the month. 
By treating each site’s time series as a function over the 744-hour interval, we apply our method to uncover patterns in the temporal variation of air pollution across locations. 
The data are also included in the R package \texttt{SLBDD}, which accompanies the book by \citet{Pena2021}. 

\subsubsection{Data preprocessing}

Before applying our method, we performed outlier removal and smoothing on the raw data. 
\begin{itemize}\itemsep=0pt
\item \textbf{Outlier removal:} 
Eight series were removed from the AirBox dataset. 
Among them, series 29 and 70 contain only a few non-zero measurements. 
The other six series are located outside Taiwan Island, while our analysis focuses on measurements within the island.
\item \textbf{Smoothing:} 
We applied a moving average filter with window size 5 to smooth the sequence of 744 measurements at each monitoring site. 
This reduces local fluctuations while preserving the overall structure relevant for clustering.
\end{itemize}

\subsubsection{Tuning parameters for clustering}

We applied the BFMS algorithm described in Section~\ref{fast_alg}, employing the truncated Gaussian kernel given in~\eqref{eq:truncated_Gaussian}.
Details of implementation settings, such as the choice of $h$ and $\tau$, are explained below:
\begin{itemize}\itemsep=0pt
\item 
The kernel bandwidth $h$ was set in the same way as \eqref{eq:bandwith_schedule}, increasing with the iteration so that smaller clusters are identified in early stages and gradually merged into larger clusters later on.

\item 
The influence range $\tau$ was set to the $25^{\rm th}$ percentile of pairwise $\Ltwo$ distances among the 508 monitoring sites, computed after outlier removal and smoothing.

\item 
Since the dataset is relatively small, we used the entire dataset without partitioning.  
The stochastic variant of the algorithm, designed to handle large-scale data, will be employed in the Argo dataset analysis presented in Section~\ref{sec:argo} to demonstrate its scalability.
\end{itemize}

%
%
\subsection{Results}

\begin{figure}[H]
\centering
\begin{minipage}[t]{0.48\textwidth}
	\centering
	\includegraphics[width=\linewidth]{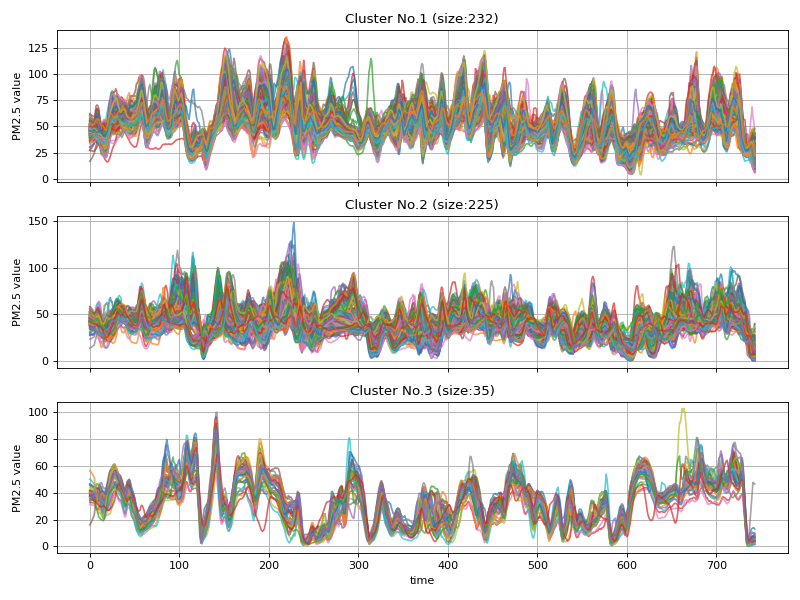}
	\caption*{(a) Hourly PM$_{2.5}$ trajectories by cluster}
\end{minipage}
\hfill
\begin{minipage}[t]{0.48\textwidth}
	\centering
	\includegraphics[width=\linewidth]{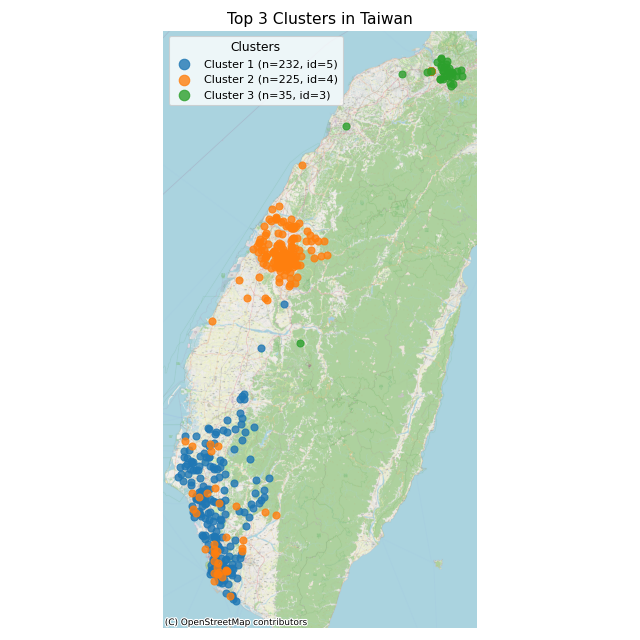}
	\caption*{(b) Geographic locations of monitoring sites, colored by cluster}
\end{minipage}
\caption{
	Clustering result based on hourly PM$_{2.5}$ trajectories from the AirBox dataset.  
	Each curve in Panel~(a) represents a monitoring site and shows its temporal PM$_{2.5}$ variation, colored by cluster.  
	Panel~(b) displays the geographic locations of the sites, also colored by cluster, illustrating the spatial distribution of each group.
}
\label{fig:airbox_results}
\end{figure}

Figure~\ref{fig:airbox_results} shows the result of clustering based on hourly PM$_{2.5}$ trajectories:
\begin{itemize}\itemsep=0pt
\item
Panel~(a) displays these trajectories, grouped into three separate boxes corresponding to the top three clusters by size.  
Each curve represents a monitoring site and illustrates the temporal variation in PM$_{2.5}$ concentrations.  
\item
Panel~(b) shows the geographic locations of the monitoring sites, colored by cluster assignment.
To improve visual clarity, only the top three clusters are shown on the map.
\end{itemize}

The top three clusters include 492 out of the 508 monitoring sites, covering nearly the entire dataset.  
This suggests that these clusters likely capture representative and interpretable temporal patterns.  
Interestingly, the clusters appear to align with the southern, middle, and northern regions of Taiwan, with some areas classified as middle overlapping with the southern region.
It is worth emphasizing that the clustering was performed solely based on the PM$_{2.5}$ trajectories; no geographic information was used.  
It is reassuring to see that the proposed mean-shift algorithm can produce results that match largely with geographical regions of the AirBoxes. 
While a detailed causal analysis is beyond the scope of this paper, as residents of Taiwan, we note that several regional factors could plausibly influence the temporal variation of PM$_{2.5}$ concentrations.  
Possible contributing factors include:
\begin{itemize}\itemsep=0pt
\item \textbf{Meteorological conditions and topography}:
Meteorological and geographical characteristics vary significantly across different regions of Taiwan, and these variations likely contribute to the observed differences in the temporal patterns of PM$_{2.5}$ concentrations.
In particular, wind speed and atmospheric stability (e.g., temperature inversions) differ between the northern, central, and southern regions.
Generally, stronger winds promote the dispersion of airborne particles, while weak or stagnant wind conditions tend to cause pollutant accumulation near the ground.
Temperature inversions, where warm air overlays cooler surface air, typically occur during nighttime or early morning and suppress vertical mixing, leading to the buildup of PM$_{2.5}$ near the surface.
Taichung, located in central Taiwan, is situated in a basin surrounded by mountains on three sides.
This topography can restrict horizontal airflow and also create favorable conditions for temperature inversions.
As a result, atmospheric stagnation becomes more likely, which in turn may lead to higher PM$_{2.5}$ concentrations and distinct temporal patterns compared to other regions.

\item \textbf{Differences in emission sources}:  
The dominant sources of PM$_{2.5}$ emissions vary by region.  
In the south, areas like Kaohsiung are home to heavy industries such as steel production, petrochemical plants, and port-related activities.  
These facilities often operate continuously, leading to relatively stable emission levels throughout the day.  
In contrast, northern cities like Taipei are characterized by dense traffic, with emission peaks typically occurring during morning and evening rush hours.  
In addition to local sources, Taiwan is also affected by long-range transport of pollutants from mainland China, particularly during the winter and early spring months.  
PM$_{2.5}$ emitted from industrial regions in China can be carried by the northeastern monsoon and elevate background concentrations across the island.  
Due to prevailing wind directions, northern Taiwan tends to be impacted earlier and more frequently.
\end{itemize}

%
%

\section{Application to Argo profiles}\label{sec:argo}

In this section, we  turn to the Argo dataset, a large-scale oceanographic collection containing temperature and salinity profiles at multiple depths, to demonstrate the scalability and broader applicability of our method.

%
%

\subsection{Dataset, preprocessing, and clustering configuration}

This section describes the Argo dataset used in our study, the preprocessing steps applied to handle its irregularly-spaced functional structure, and the experimental setup adopted for clustering analysis.

\subsubsection{Dataset}

The Argo program is an international project jointly supported by many countries and has been systematically collecting temperature and salinity profiles from autonomous profiling floats deployed across the global oceans since the year 2000. 
(See \url{https://argo.ucsd.edu/} for details.)
These floats drift with ocean currents and periodically dive to depths corresponding to pressures of up to approximately 2000 decibars, recording vertical profiles of temperature and salinity before resurfacing and transmitting the collected data via satellite.
As a result, Argo provides a vast dataset that enables large-scale oceanographic and climate studies \citep{wong2020argo}.

In this study, we used Argo data collected between 2006 and 2016, analyzing approximately one million profiling cycles ($n \approx 10^6$).
Each profile consists of temperature and salinity measurements taken at different pressure levels (up to $p \approx 2000$).
However, these measurements are not uniformly sampled at fixed depth levels; rather, the pressure values vary across profiles, resulting in an irregularly sampled functional data structure.
Functional data approaches have previously been applied to Argo profiles to extract meaningful oceanographic patterns; see, for example, \citet{yarger2022functional}.

\subsubsection{Data preprocessing}
The preprocessing steps are summarized below.
\begin{itemize}
\item
{\bf Profiling cycle construction and initial filtering:}
Each individual measurement in the Argo dataset is associated with metadata, including platform number, observation time, geographic location, and physical variables such as pressure, temperature, and salinity.
Measurements that share the same platform number, observation time, and location are grouped into a single profiling cycle, consisting of multiple measurements taken at different pressure levels during one dive by a float.
As part of our preprocessing, we excluded any cycles containing fewer than 20 valid data points to ensure sufficient vertical resolution and data quality.

\item
{\bf Interpolation and selection of analysis-ready cycles:}
Since temperature and salinity measurements are recorded at irregular pressure levels, we applied cubic spline interpolation to each profiling cycle in order to map the data onto a common set of grid points spanning the pressure range $[0, 2000]$.
However, interpolated values near the boundaries (i.e., around $p = 0$ and $p = 2000$) tend to be unreliable due to the scarcity of observed data points in those regions.
To avoid introducing artifacts through extrapolation, we set the interpolated values to NaN (denoting missing values) wherever extrapolation would be required.
For subsequent analysis, we restricted the pressure domain to the range $[20, 300]$.
Furthermore, only profiling cycles that contained no missing values within this interval were retained for downstream clustering analysis.
After this filtering step, the resulting dataset consisted of slightly over one million valid profiling cycles ($n = 1{,}024{,}852$).
\end{itemize}

\subsubsection{Tuning parameters for clustering}

We applied the stochastic fast BFMS algorithm introduced in Section~\ref{fast_alg}, using the truncated Gaussian kernel given in~\eqref{eq:truncated_Gaussian}.
The choices of parameters, including $h$, $\tau$  in~\eqref{eq:truncated_Gaussian}, and the number of partitions, are explained below:
\begin{itemize}\itemsep=0pt
\item 
The kernel bandwidth $h$ followed the schedule in \eqref{eq:bandwith_schedule}, where $h$ increases with the iteration.
\item 
The influence range $\tau$ was determined as the $25^\th$ percentile of pairwise $\Ltwo$ distances computed from 5,000 randomly sampled profiling cycles, separately for temperature and salinity.
\item
To ensure computational scalability, the full set of profiling cycles was randomly partitioned at each iteration into $1,024$ disjoint subsets, each containing approximately $1,000$ samples.
\end{itemize}

%
%
\subsection{Results}

\begin{figure}[H]
\centering
\begin{subfigure}[t]{0.48\linewidth}
	\centering
	\includegraphics[width=\linewidth]{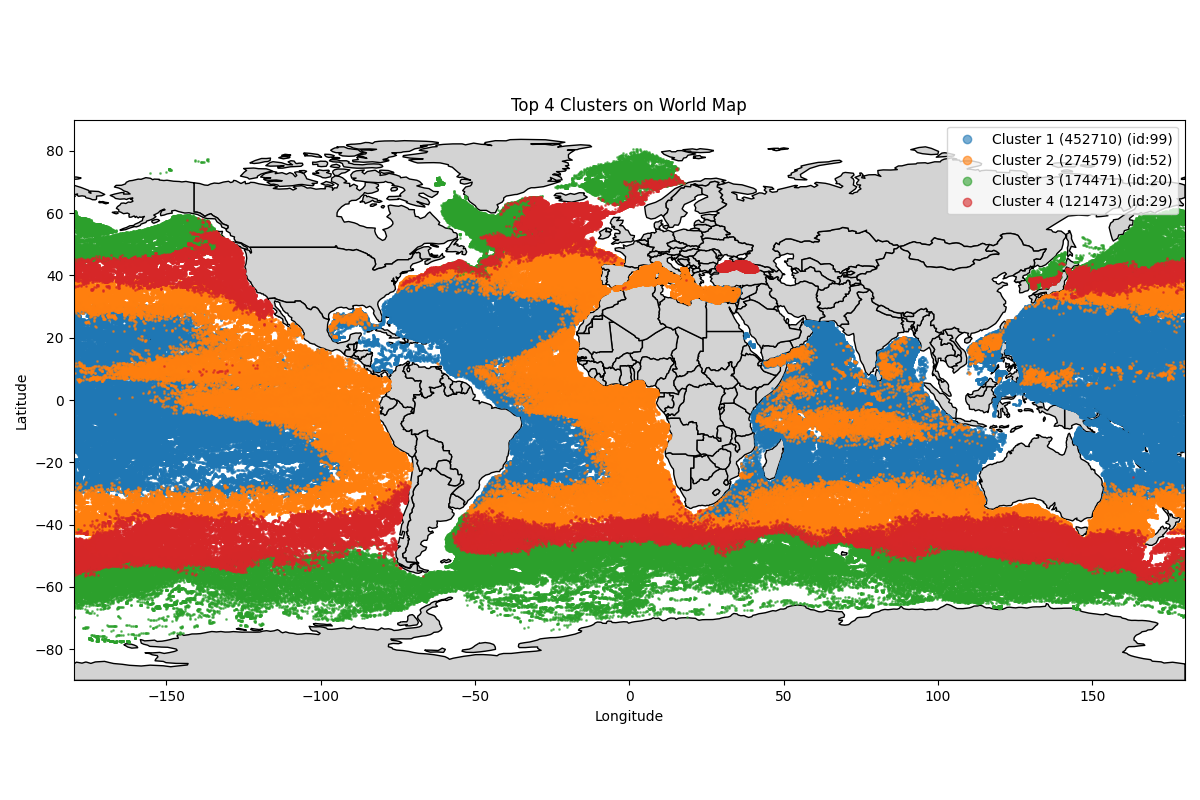}
	\caption{aggregated}
\end{subfigure}
\hfill
\begin{subfigure}[t]{0.48\linewidth}
	\centering
	\includegraphics[width=\linewidth]{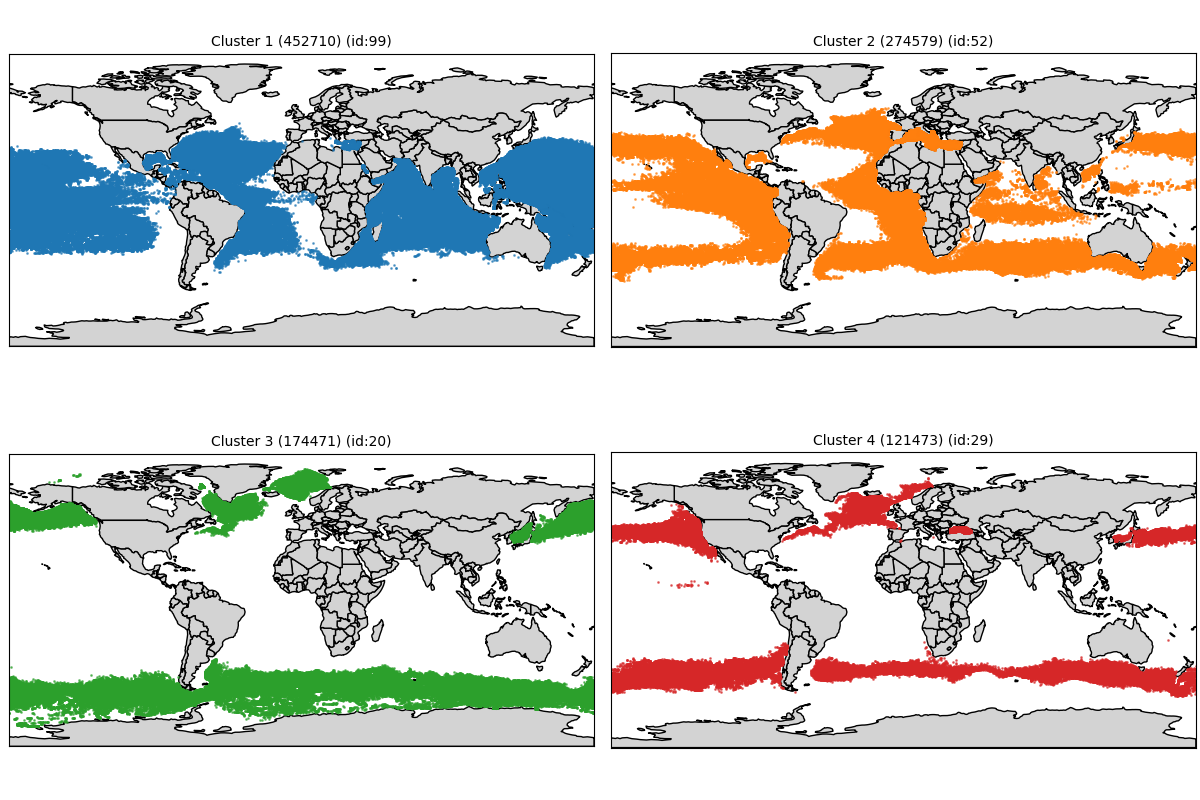}
	\caption{separated}
\end{subfigure}
\caption{Cluster maps of temperature profiles: (a) the four largest clusters overlaid on a single map, (b) the same clusters shown separately.}
\label{fig:temperature_map}
\end{figure}

\begin{figure}[htbp]
\centering
\begin{subfigure}[t]{0.48\linewidth}
	\centering
	\includegraphics[width=\linewidth]{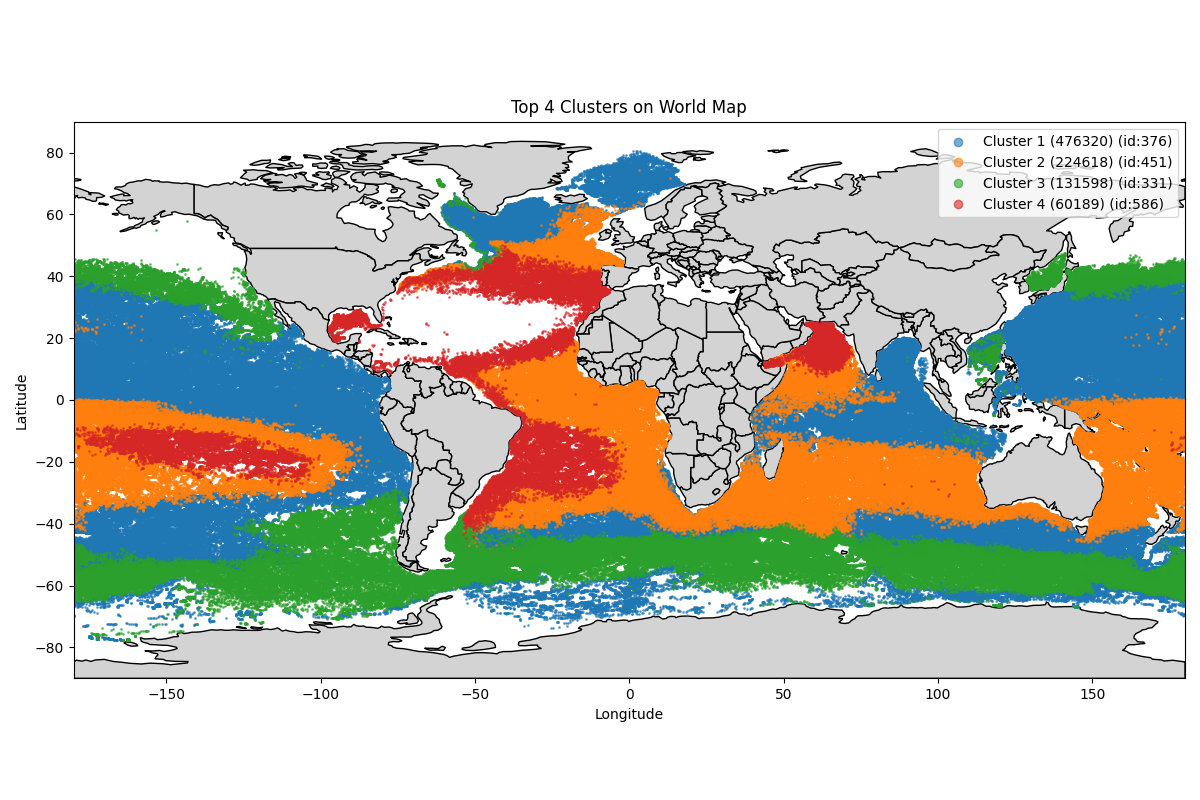}
	\caption{aggregated}
\end{subfigure}
\hfill
\begin{subfigure}[t]{0.48\linewidth}
	\centering
	\includegraphics[width=\linewidth]{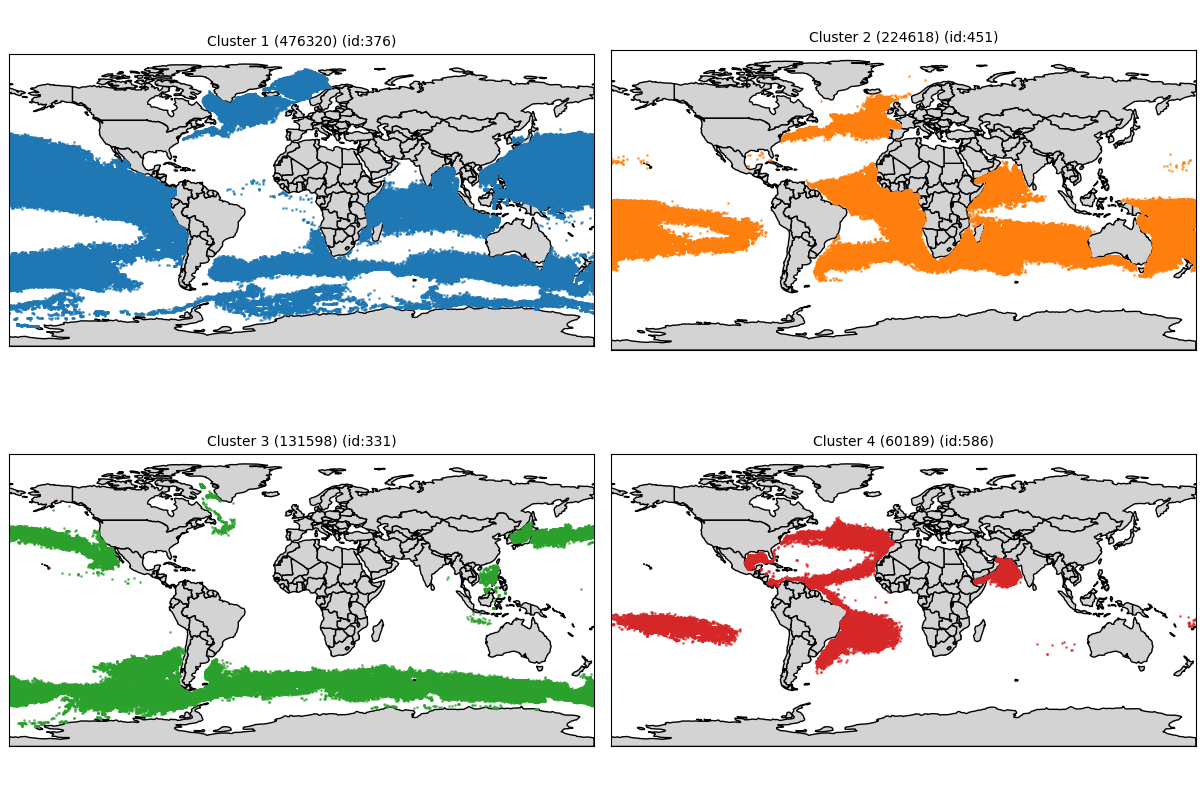}
	\caption{separated}
\end{subfigure}
\caption{Cluster maps of salinity profiles: (a) the four largest clusters overlaid on a single map, (b) the same clusters shown separately.}
\label{fig:salinity_map}
\end{figure}

Figures~\ref{fig:temperature_map} and \ref{fig:salinity_map} visualize the clustering results for temperature and salinity data, respectively, over geographic coordinates. In each case, only the profiling cycles belonging to the four largest clusters are shown, with each cluster indicated by a different color. 
Profiling cycles associated with smaller clusters are omitted from the figures. 
In both figures, Panel~(a) displays all four clusters overlaid on a single map, whereas Panel~(b) shows them separately for ease of interpretation.

Importantly, geographic location information was not used in the clustering process; only the temperature and salinity profiles were provided as input. 
Nevertheless, the resulting clusters appear to correspond roughly to geographically distinct regions. 
We also observe that the spatial distribution of some clusters bears certain resemblance to known ocean current patterns, illustrating the usefulness of proposed method in the absence of domain knowledge.
Notably, some clusters align with known features such as high-latitude regions, suggesting that certain oceanographic structures may indeed be reflected in the clustering patterns.


This example clearly demonstrates that our algorithm can handle very large datasets within practical computational time and memory limits , in accordance with its lower complexity established in Section~\ref{complexity}.
\footnote{Approximately 5--6 hours for the two-round clustering of temperature and salinity combined, on a Linux server with two AMD EPYC 7742 64-core processors and 1~TB of RAM.}
By contrast, executing the full-data mean shift algorithm without random partitions would require computing and storing the complete pairwise distance matrix at each iteration, which for $n \approx 10^6$ would demand approximately 4~TB of RAM, making it infeasible on typical hardware.


%
%

\section{Conclusion}\label{sec:conclusion}

We have proposed a blurring functional mean shift algorithm for functional data in a Hilbert space.
To address the computational challenges posed by large-scale functional datasets, we developed a stochastic variant based on random data partitioning that substantially reduces computational cost.
We also established a rigorous convergence analysis for the full-data blurring functional mean shift procedure, which provides a theoretical foundation for its practical use.
While a complete convergence theory for the stochastic variant remains an open problem, we derived a functional law of large numbers result that provides partial theoretical justification for its use.
The proposed stochastic variant was applied to Argo oceanographic profiles, illustrating its scalability and practical usefulness in real-world functional data analysis.

These results suggest several directions for future research, including adapting the framework to other types of functional or structured data, exploring alternative partitioning strategies for stochastic updates, and extending the algorithm to broader application domains. In particular, a more complete theoretical understanding of the convergence properties of the stochastic variant remains an important topic for future work.

%
%

\section*{Acknowledgements}



This research was partially supported by the National Science and Technology Council, Taiwan (Grant IDs: NSTC 113-2118-M-001 -015 -MY2 and NSTC 113-2118-M-001 -016 -MY2). 
The authors declare that they have no competing interests.

\section*{Disclosure Statement}

No competing interest is declared.

\section*{Data Availability Statement}
The data that support the findings of this study are derived from publicly available sources.

The Taiwan PM2.5 data are from the AirBox dataset \citep{Chen2017_PM25}, which are available through the R package SLBDD associated with the book by \citet{Pena2021}.

The Argo oceanographic data are available from the Argo Global Data Assembly Centers (GDAC) at https://argo.ucsd.edu and https://www.ifremer.fr/argo. Access to these data may require registration and is subject to the data usage policies of the Argo program.



%
%

\appendix

\vskip 0.2in
\bibliography{references}

\newwrite\cntfile
\immediate\openout\cntfile=eqcount.tex
\immediate\write\cntfile{\string\setcounter{equation}{\number\value{equation}}}
\immediate\write\cntfile{\string\setcounter{figure}{\number\value{figure}}}
\immediate\write\cntfile{\string\setcounter{table}{\number\value{table}}}
\immediate\closeout\cntfile

\end{document}